\documentclass[12pt]{article}
\usepackage{amsmath}
\usepackage{graphicx}
\usepackage{natbib}
\usepackage{url} 
\usepackage{amssymb, bbm}
\usepackage{multirow, subfig}
\usepackage{amsthm}
\usepackage{bbold}
\usepackage{algorithm}
\usepackage{algpseudocode}
\algnewcommand\algorithmicinput{\textbf{INPUT:}}
\algnewcommand\INPUT{\item[\algorithmicinput]}
\algnewcommand\algorithmicoutput{\textbf{OUTPUT:}}
\algnewcommand\OUTPUT{\item[\algorithmicoutput]}

\usepackage{tikz}
\usetikzlibrary{shapes}
\tikzstyle{cblue}=[circle, draw, thin, fill=cyan!20, scale=0.8]
\tikzstyle{qgre}=[rectangle, draw, thin, fill=green!20, scale=0.8]

\usepackage{hyperref}[]
\hypersetup{
    colorlinks=true,
    linkcolor=blue,
    filecolor=magenta,      
    urlcolor=cyan,
      citecolor=blue,
    }

\usepackage{cleveref}

\newtheorem{definition}{Definition}

\DeclareMathOperator*{\argmin}{argmin}
\DeclareMathOperator*{\argmax}{argmax}

\newcommand{\blind}{0}

\begin{document}

\def\spacingset#1{\renewcommand{\baselinestretch}%
{#1}\small\normalsize} \spacingset{1}


\if0\blind
{
  \title{\bf Graph matching beyond \\
perfectly-overlapping Erd\H{o}s--R\'enyi random graphs}
  \author{Yaofang Hu, \, Wanjie Wang \\
   Department of Statistics, National University of Singapore\\
    and \\
	Yi Yu\\
    Department of Statistics, University of Warwick}
  \maketitle
} \fi

\if1\blind
{
  \bigskip
  \bigskip
  \bigskip
  \begin{center}
    {\LARGE\bf Title}
\end{center}
  \medskip
} \fi

\bigskip
\begin{abstract}
Graph matching is a fruitful area in terms of both algorithms and theories.  In this paper, we exploit the degree information, which was previously used only in noiseless graphs and perfectly-overlapping Erd\H{o}s--R\'enyi random graphs matching.  We are concerned with graph matching of partially-overlapping graphs and stochastic block models, which are more useful in tackling real-life problems.  We propose the edge exploited degree profile graph matching method and two refined varations.  We conduct a thorough analysis of our proposed methods' performances in a range of challenging scenarios, including a zebrafish neuron activity data set and a coauthorship data set.  Our methods are proved to be numerically superior than the state-of-the-art methods.
\end{abstract}

\noindent%
{\it Keywords:}  	Graph matching; Degree profile; Partially-overlapping graphs correlated Bernoulli networks; Stochastic block models.


\section{Introduction}
Graph matching has been an active area of research for decades. The research on graph matching can be traced back to at least 1970s \citep[e.g.][]{ullmann1976algorithm}, and interpreted as ``graph matching'', ``network alignment'' and ``graph isomorphism''. In this paper, we do not distinguish these terms, nor the terms ``graph'' and ``networks'', or ``nodes'' and ``vertices''.   Mathematically, the graph matching problem can be loosely stated as follows. Given two graphs $G_1 = (V_1, E_1)$ and $G_2 = (V_2, E_2)$, it is assumed that $V_1$ and $V_2$ are the same or largely overlapped upon an unknown permutation $\pi^*$.  Graph matching is to seek the mapping $\pi^*$ between the vertices sets $V_1$ and $V_2$.  A correct matching would help augment the connectivity information between the vertices, and hence improve the graph analysis.  In recent years, due to the advancements in collecting, storing and processing large volume of data, graph matching is going through a renaissance, with a surge of work on graph matching in different application areas.  For instance, \cite{narayanan2009anonymizing} targeted at acquiring information from an anonymous graph of Twitter with the graph of Flickr as the auxiliary information; \cite{kazemi2016proper} seek the alignment of protein-protein interaction networks in order to uncover the relationships between different species; \citep{haghighi2005robust} constructed graphs based on texts relationship and developed a system for deciding whether a given sentence can be inferred from text by matching graphs.

Graph matching is an extremely fruitful research area.  In the following, we review the existing literature from three different aspects, based on which, we characterize our main interest of this paper.

In terms of methodology, broadly speaking, the graph matching algorithms can be categorized into two schools: exact matching and inexact matching.  The exact graph matching focus on deterministic graphs. It seeks a perfect matching, which is NP-hard in most cases, with exceptions in some special graph structures, for instance planar graphs \citep[e.g.][]{eppstein2002subgraph}.  When we move from deterministic graphs to random graphs, it is challenging and not natural to seek a perfect matching. The inexact matching approaches are considered in this case.  Existing methods designed for inexact matching include tree search types of methods \citep[e.g.][]{sanfeliu1983distance}, continuous optimization types of methods \citep[e.g.][]{liu2012extended} and spectral-based convex relaxation types of methods.  Due to the demand of computational feasibility when dealing with large-scale datasets, the spectral-based methods have been, arguably, the most popular type of methods.  To be more specific, spectral-based methods include spectral matching \citep[e.g.][]{leordeanu2005spectral}, semidefinite-programming approaches \citep[e.g.][]{schellewald2005probabilistic} and doubly-stochastic relaxation methods \citep[e.g.][]{gold1996graduated}.  For comprehensive reviews, we refer to \cite{conte2004thirty}, \cite{foggia2014graph} and \cite{yan2016short}.

In terms of the underlying models, despite the large amount of algorithms proposed over the years, the majority of the efforts are on the Erd\H{o}s--R\'enyi random graphs \citep{erdos1959random}, which are fundamental yet realistic.   Beyond the Erd\H{o}s--R\'enyi random graphs, \cite{patsolic2017vertex} studied the graph matching in a random dot product graph \citep{young2007random} framework.  \cite{li2016matching} is concerned with the community matching in a multi-layer graph, the matching resolution thereof is at the community level, but not at the individual level.   The study in this area usually is complicated by the misclustered vertices. 

In terms of the proportion of overlapping vertices in two graphs. Some of the existing works consider the situations where the two graphs have identical vertices sets, while some consider the situations where the difference between two vertices sets is nonempty.  In the sequel, we will refer to these two situations as \emph{perfectly-overlapping} and \emph{partially-overlapping}.  Work on the latter includes the following: \cite{pedarsani2011privacy} studied the privacy of anonyized networks; \cite{kazemi2015can} defined a cost function for structural mismatch under a particular alignment and established a threshold for perfect matchability; and \cite{patsolic2017vertex} provided a vector of probabilities of possible matchings.

We now specify the problem we are concerned about in this paper.  (1) We intend to exploit the degree information and extend the degree profile method, which shares connection with the doubly stochastic relaxation methods and which has been previously studied in  \cite{czajka2008improved} and \cite{mossel2017shotgun} for deterministic graphs and in \cite{ding2018efficient} for perfectly-overlapping Erd\H{o}s--R\'enyi random graphs. (2) We consider network models with community structures, including stochastic block models, which is arguably the most popular network models for both theoretical and practical studies. (3) We tackle partially-overlapping graphs, e.g.~the two graphs to be matched do not have identical vertices sets.    Our contribution is listed below.
	\begin{itemize}
	\item We formally describe a partially-overlapping correlated Bernoulli networks model in \Cref{def-corr-bern-net}. Further, we explore the degree profile graph matching method for the newly defined partially-overlapping Erd\H{o}s--R\'enyi random graphs and also the stochastic block random graphs.  To the best of our knowledge, this is the first work to exploit the degree profile-type method in stochastic block model graph matching problems.
	\item We propose the edge exploited (EE) degree profile graph matching method.  In addition, we propose refined EE algorithms, including pre-processing and post-processing steps.  These proposed methods are demonstrated to outperform the state-of-the-art methods when the graphs are partially overlapping.
	\item The degree profile core of our methods enable us to conduct graph matching in the sparse regime, where the spectral-based methods usually fail.
	\end{itemize}

The rest of this paper is organized as follows.  In \Cref{sec-methods}, we present our proposed methods.  We kick off by reviewing a state-of-the-art degree profile method, extend it to handle the partially-overlapping scenarios, and finally tackle stochastic block model graph matching.  Our proposed methods are supported by extensive numerical evidence on both simulated and real datasets in Sections \ref{sec-numerical}-\ref{sec-data}.  The paper is concluded by discussions in \Cref{sec-diss}.

\section{Methodology}\label{sec-methods}

In this section, we first state the partially-overlapping correlated Bernoulli models in \Cref{sec-setup}, and introduce the degree profile graph matching method in \Cref{sec-dp}.  In \Cref{sec-EE}, we propose the core edge exploited (EE) graph matching method, with its refinements in \Cref{sec-refine}.  The stochastic block models graph matching is tackled in \Cref{sec-comm}.

\subsection{Correlated Bernoulli networks}\label{sec-setup}

The degree profile method was pioneered in \cite{czajka2008improved} and \cite{mossel2017shotgun} on graph matching of two identical graphs generated from Erd\H{o}s--R\'enyi random graph models.  This method is further studied in \cite{ding2018efficient} and is extended to correlated Erd\H{o}s--R\'enyi random graphs.  The key of the degree profile graph matching is to assign each vertex an empirical distribution of its neighbours' degrees, and match vertices by measuring the distance between each pair of the empirical distributions.  We first set up the models in this section.

\begin{definition}[Bernoulli networks $\mathcal{G}(n, \Theta)$] \label{def-bern-net} 
A network with vertices set $\{1, \ldots, n\}$ is a Bernoulli network $\mathcal{G}(n, \Theta_{n \times n})$, if its associated adjacency matrix $A \in \mathbb{R}^{n \times n}$, which is defined by $A_{ii} = 0$, $i \in \{1, \ldots, n\}$,
	\[
		A_{ij} = A_{ji} = \begin{cases}
 			1, & \mbox{vertices $i$ and $j$ are connected by an edge,}\\
 			0, & \mbox{otherwise},
 		\end{cases}
	\]
	where $\{A_{ij}, \, i < j\}$ are independent Bernoulli random variables with $\mathbb{E}(A) = \Theta$.	  
\end{definition}

\Cref{def-bern-net} includes the Erd\H{o}s--R\'enyi random graphs, where all the off-diagonal entries of $\Theta$ are equal; stochastic block models, where $\Theta$ possesses a block structure; degree corrected block models, where degree heterogeneity is added; random dot product graphs, where latent positions are assumed.  Note that in \Cref{def-bern-net}, it is assumed that matrices $A$ and $\Theta$ are symmetric with diagonal entries being zero.  In fact, the definition along with the methods proposed later in this paper can be relaxed to more general cases.  However, in this paper, we focus on \Cref{def-bern-net} and move on to more general cases in \Cref{sec-diss}.

\begin{definition}[Partially-overlapping correlated Bernoulli networks] 
\label{def-corr-bern-net}  
	Let $G$ be the adjacency matrix of a given graph and $s, \rho \in [0, 1]$ be the overlapping and correlation parameters, respectively.  Construct a matrix $A'$ by independently keeping or removing each row (and the corresponding column) in $G$ with probability $1 - s$. Further, construct $A$  by $A_{ji} = A_{ij} = A'_{ij}  X_{ij}$ where $X_{ij} \stackrel{i.i.d.}{\sim} Bernoulli(\rho)$, $i < j$. The graph with adjacency matrix $A$ is called a child graph of $G$. 
 Relabel the vertices of $G$ according to a latent permutation $\pi^*$ and then repeat the sampling process independently to obtain another child graph $B$. $A$ and $B$ are partially-overlapping correlated Bernoulli networks. 
\end{definition}

We list a few cases to better understand \Cref{def-corr-bern-net}.  When $s = \rho = 1$, $A$ and $B$ are exactly the same up to a permutation \citep[isomorphic graphs, e.g.][]{scheinerman2011fractional}.  If we fix $s = 1$ only, and let $G$ be a realization of $\mathcal{G}(n, \Theta)$ in \Cref{def-bern-net}, where $\Theta$ has off-diagonals as a constant, then both $A$ and $B$ have all $n$ vertices.  In this case, $A$ is in fact an adjacency matrix of $\mathcal{G}(n, s\Theta)$, and $B$ can be seen as 
	\[
		B_{\pi^*(i)\pi^*(j)} \sim \begin{cases}
 			\mathrm{Ber}(\rho), & A_{ij} = 1, \\
 			\mathrm{Ber}\left(\frac{s\Theta_{ij} (1-\rho)}{1 - s\Theta_{ij}}\right), & A_{ij} = 0.
 		\end{cases}
	\]
 Hence, it coincides with the perfectly-overlapping correlated Erd\H{o}s--R\'enyi random graphs, which have been studied extensively in the existing literature including \cite{lyzinski2014seeded} and \cite{ding2018efficient}, among others.  When $\rho = 1$, both $A$ and $B$ are subgraphs of $G$.  

Compared to the perfectly-overlapping correlated Erd\H{o}s--R\'enyi random graphs, \Cref{def-corr-bern-net} characterizes a general model, but inherits the key features that (i) $A$ and $B$ have identical marginal distributions, and (ii) the corresponding entries of $A$ and $B$ are correlated with correlation $\rho$.  In practice, the underlying $G$ is usually unknown, but $A$ and $B$ can be obtained from different studies.  For instance, one may obtain a fully-known Amazon users network and an anonymized eBay users network, while the underlying true network is unknown.  
Due to the anonymity, it is only reasonable to assume the users are largely overlapping in these two networks, but not perfectly. 

The goal of this paper is to match the vertices between $A$ and $B$. Since exact matching between $A$ and $B$ may not exist, we seek best matching between largest overlapped subgraphs of $A$ and $B$.  Mathematically, for networks $A$ and $B$ with $n_A$ and $n_B$ vertices, respectively, we seek a permutation $\widehat{\pi}$ defined as
	\[
		\widehat{\pi} \in \argmax_{\Pi_m} \max_{1\leq m \leq \min\{n_A, n_B\}} \max_{S_A: |S_A| = m, S_A \subset [n_A]} \max_{S_B: |S_B| = m, S_B \subset [n_B]}\langle A_{S_A}, \Pi B_{S_B} \Pi^{\top}\rangle,
	\]  
	where $\Pi_m$ ranges over all $m \times m$ permutation matrices and $\langle \cdot, \cdot \rangle$ denotes the matrix inner product. 

\subsection{Degree profile graph matching}\label{sec-dp}

Generally speaking, degree profile graph matching methods exploit the degree information of all the neighbours to construct an empirical distribution for each vertex, and then match the vertices by comparing the similarity between these empirical distributions.   In this section, we first detail the definition of degree profile and then explain the simplest form of the degree profile method in \Cref{alg-graph-matching}. 

\begin{definition}[Degree profile]\label{def-degree-profile}
Let $A \in \mathbb{R}^{n \times n}$ be an adjacency matrix.  For any $i \in \{1, \ldots, n\}$, let $a_i = \sum_{j = 1}^n A_{ij}$ and $N_A(i) = \{j: \, A_{ij} = 1\}$ be the degree and the neighbourhood of $i$, respectively.  Further denote $a_k^{(i)}$ as the degree of $i$'s neighbour $k$ that $a_k^{(i)} = \sum_{l = 1}^n A_{lk}$.  Let $\mu_i(x) = a_i^{-1} \sum_{k \in N_A(i)} \mathbbm{1}\{a_k^{(i)} \leq x\}$, $x \in \mathbb{R}$, be the empirical cumulative distribution functions of the set $\{a_k^{(i)}, \, k \in N_A(i)\}$. The degree profile of vertex $i$ in $A$ is defined to be $\mu_i(\cdot)$ and denoted as $\mathrm{DP}(A, i)$.	
\end{definition}

The degree profile defined in \Cref{def-degree-profile} is the second term of the iterated degree sequence.  A necessary and sufficient condition for fractional isomorphism is that two graphs have identical iterated degree sequences.  See \cite{scheinerman2011fractional} for more details.

In \cite{ding2018efficient}, a similar definition is studied for the perfectly-overlapping Erd\H{o}s--R\'enyi random graphs. In \cite{ding2018efficient}, the degree profile is a normalized empirical distribution of neighbours' degrees, excluding edges between neighbours when counting degrees and standardizing the degrees such that they are mean zero and variance one random variables.  This normalization is for theoretical simplicity when dealing with the behaviours of the empirical distributions. 

With the degree profiles for all vertices in $A$ and $B$, our next step is to introduce a distance (or similarity) between each pair $(i, j)$, $i \in A$ and $j \in B$ and construct an $n_A \times n_B$ distance matrix $W$ (or similarity matrix), where $W_{ij}$ denotes the distance (or similarity) between the degree profiles of $i$ and $j$.   Intuitively, if $i \in A$ and $j \in B$ is a true pair, that is to say $\pi^*(i) = j$, then the distance $W_{ij}$ is small (or the similarity $W_{ij}$ is large), otherwise large (small).  For each $i \in A$, hence, we seek the mapping $\widehat{\pi}(i) = \argmin_{j \in B} W_{ij}$ (or $\widehat{\pi}(i) = \argmax_{j \in B} W_{ij}$).

The mapping $\hat{\pi}$ may not be a permutation, since multiple vertices in $A$ might be mapped to the same vertex $j$ in $B$.  Hence, the final graph matching is output by applying a maximal bipartite matching algorithm to this mapping $\hat{\pi}$.  Every vertex in $B$ is matched to at most one vertex in $A$ and there is no guarantee that all the vertices in $A$ are matched to vertices in $B$.  It is ensured that there exists no other bipartite matching which can match more vertices.  In this paper, this is done by R \citep{R} package igraph \citep{Rigraph}, which uses the push-relabel algorithm introduced in \cite{cherkassky1998augment}.

With the analysis, we detail the, arguably, simplest degree profile algorithm, which invloves a subroutine of calculating the distance matrix in \Cref{alg-0} and implements the degree profile graph matching in \Cref{alg-graph-matching}. 

\begin{algorithm}[!htbp]
	\begin{algorithmic}
		\INPUT $A \in \{0, 1\}^{n_A \times n_A}$, $B \in \{0, 1\}^{n_B \times n_B}$
		\For{$i = 1, \ldots, n_A$}
			\State $\mu_i \leftarrow \mathrm{DP}(A, i)$ 
		\EndFor
		\For{$i = 1, \ldots, n_B$}
			\State $\nu_i \leftarrow \mathrm{DP}(B, i)$ 
		\EndFor
		\For{$i = 1, \ldots, n_A$}
			\For{$j = 1, \ldots, n_B$}
				\State $W_{ij} \leftarrow \mathrm{W}_1(\mu_i, \nu_j)$
			\EndFor 
		\EndFor
		\OUTPUT $W$
		\caption{Constructing distance matrix $W(A, B)$ \label{alg-0}}
	\end{algorithmic}
\end{algorithm}

\begin{algorithm}[!htbp]
	\begin{algorithmic}
		\INPUT $A \in \{0, 1\}^{n_A \times n_A}$, $B \in \{0, 1\}^{n_B \times n_B}$
		\State $W \leftarrow W(A, B)$
		\State $Z \leftarrow 0_{n_A \times n_B}$, which is an all zero matrix
		\For{$i = 1, \ldots, n_A$}
			\State $k \leftarrow \argmin_{j = 1}^{n_B} W_{ij}$
			\State $Z_{ik} \leftarrow 1$
		\EndFor
		\State $\widehat{\pi} \leftarrow \mathrm{MaxBipartiteMatching}(Z)$
		\OUTPUT $\widehat{\pi}$
		\caption{Degree profile graph matching \label{alg-graph-matching}}
	\end{algorithmic}
\end{algorithm}

In \Cref{alg-0}, the distance used thereof is $W_1$, the $1$-Wasserstein distance \citep[e.g.][]{villani2009wasserstein}.  In fact, any distance or similarity measure can be used here.  We choose $W_1$ to demonstrate numerical results and it will be used throughout this paper.   In \Cref{sec-refine}, in the refined algorithms that \Cref{alg-edge-exp-dp-pre} and \ref{alg-edge-exp-dp-post}, the similarity measure used is the number of common neighbours of $i$ and $j$ according to the prior information, which delivers satisfactory results. 

\Cref{alg-graph-matching} is similar to Algorithm 1 proposed in \cite{ding2018efficient}. A difference is that \Cref{alg-graph-matching} seeks the pair with minimal distance for each vertex, but \cite{ding2018efficient} consider $n$ pairs with smallest distances among all possible combinations. The two approaches deliver the same result if two networks are perfectly overlapping, yet \Cref{alg-graph-matching} can also provide possible matchings for vertices without counterparts.  More discussions on this are available later.

The theoretical properties of degree profile graph matching have been extensively studied in \cite{ding2018efficient}.  The main advantage of degree profile graph matching over competitors is the ability to conduct polynomial-time graph matching in a sparse regime, where the spectral-type methods would fail. The main challenges in deriving the theoretical properties of the output of \Cref{alg-graph-matching} is to carefully control the fact that the degree profiles are linear combinations of correlated random variables, and this is out of the scope of this paper.  

\subsection{Edge exploited methods for partially-overlapping graphs}\label{sec-EE}

The state-of-the-art methodology on the degree profile graph matching method is restricted to the cases where a bijection exists between the vertices sets of two graphs.  This, however, is by no means realistic in more real-life problem. It is, therefore, of great interest to extend \Cref{alg-graph-matching} to handle the partially-overlapping networks.

\begin{algorithm}[htbp]
\begin{algorithmic}
	\INPUT $A \in \{0, 1\}^{n_A \times n_A}$, $B \in \{0, 1\}^{n_B \times n_B}$, a positive integer $d \geq 1$
	\State $W \leftarrow W(A, B)$	\Comment{\Cref{alg-0}}
	\State $Z \leftarrow 0_{n_A \times n_B}$, which is an all zero matrix
	\For{$i = 1, \ldots, n_A$}
		\State $\{(i, i_k)\}_{k = 1}^d \leftarrow $ the indices of the $d$ smallest entries in the $i$th row of $W$
		\State $(Z_{i i_k}, \, k = 1, \ldots, d)^{\top} \leftarrow (1, \ldots, 1)^{\top} \in \mathbb{R}^d$
	\EndFor
	\OUTPUT $Z$
	\caption{Edge exploited degree profile graph matching. $\mathrm{EE}(A, B, d)$\label{alg-edge-exp-dp}}
\end{algorithmic}
\end{algorithm}

\begin{figure}[!htbp]
\begin{center}
\begin{tikzpicture}
	\draw [rotate around ={30:(4.5, -0.5)}, color = blue!50, thick] (4.5, -0.5) ellipse (2cm and 1.2cm);
	\draw [rotate around ={140:(4.5, -1)}, color = green!50, thick] (4.5, -1) ellipse (2cm and 1.2cm);
	\draw [rotate around ={140:(4.5, -1)}, color = red!50, thick] (4, -1.8) ellipse (2cm and 1.2cm);
	\node[circle, fill = blue!50] (A1) at (0, 0) {};
	\node[rectangle, fill = red!50] (A2) at (-1, -1.73) {};
	\node[diamond, fill = green!50] (A3) at (1, -1.73) {};	
	\draw[thick] (A1) -- (A2);
	\draw[thick] (A2) -- (A3);	
	\draw[thick] (A1) -- (A3);	
	\node at (0, -0.9) {A};	

	\node[circle, fill = blue!50] (B1) at (4, 0) {};
	\node[diamond, fill = green!50] (B2) at (4, -1.73) {};
	\node[regular polygon,regular polygon sides=5, fill = purple!50] (B3) at (6, -1.73) {};	
	\node[regular polygon,regular polygon sides=5, fill = purple!50] (B4) at (6, 0) {};	
	\draw[thick] (B1) -- (B2);
	\draw[thick] (B2) -- (B3);	
	\draw[thick] (B1) -- (B4);		
	\draw[thick] (B3) -- (B4);	
	\node at (5, -0.9) {B};		
	
	\node (C1) at (5.5, 0.9) {};
	\node (C2) at (4.5, -2.3) {};
	\node (C3) at (6.5, -2.3) {};
	
	\draw [->, thick, draw = gray!50] (A1) to [out = 20, in = 170] (C1); 
	\draw [->, thick, draw = gray!50] (A3) to [out = -20, in = 190] (C2); 	
	\draw [->, thick, draw = gray!50] (A2) to [out = -20, in = 190] (C3); 	
\end{tikzpicture}
\caption{A cartoon of the edge exploited matching \label{fig-edge-exp}}	
\end{center}
\end{figure}
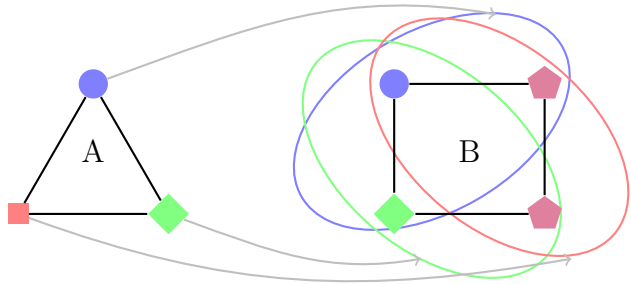

There are two differences between Algorithms~\ref{alg-graph-matching} and \ref{alg-edge-exp-dp}.  First, in \Cref{alg-edge-exp-dp}, we introduce an additional parameter $d$, which is in fact taken to be 1 in \Cref{alg-graph-matching}.  In \Cref{alg-edge-exp-dp}, the matrix $Z$ is an adjacency matrix of a bipartite graph, where each vertex in $A$ is connected to one and only one vertex in $B$.  In the edge exploited version \Cref{alg-edge-exp-dp}, we allow for $d$ edges for each vertex in $A$.  A demonstration is depicted in \Cref{fig-edge-exp} with $d = 3$.  Instead of matching each vertex in $A$ to an individual vertex in $B$, we match $A$ to a hypergraph built upon $B$ with hyper-edges of size at most $d = 3$.  Second, the final output of \Cref{alg-graph-matching} from a maximum bipartite graph matching algorithm, and it does not allow for matching one vertex to a collection of vertices.  To overcome this, we omit the maximum bipartite graph matching step in \Cref{alg-edge-exp-dp} and output the matching matrix $Z$ directly.

\begin{figure}[!htbp]
\begin{center}
	\subfloat{\includegraphics[width = 0.5\textwidth]{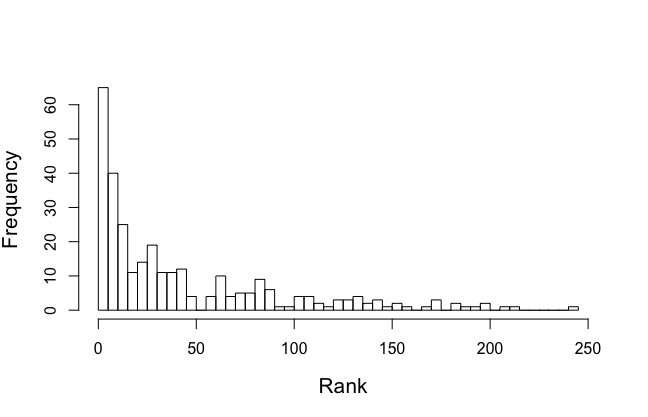}}	
	\subfloat{\includegraphics[width = 0.5\textwidth]{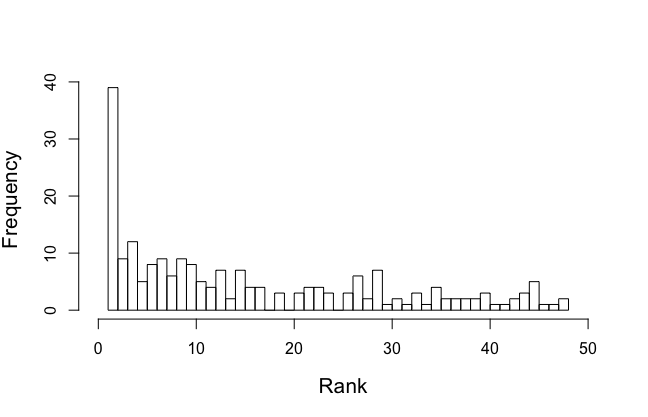}}	
	\caption{Motivation for the EE algorithm.  The $x$-axis is for the rank and $y$-axis is for the frequency. The left panel is the histogram of the ranks the distance between true pairs among all pairs. The right panel is a zoomed-in version of the top 50 ranks in the left panel.  \label{fig-EE-motive}}
\end{center}
\end{figure}

To see why it is necessary to consider matching a node with more than one nodes, we match two partially-overlapping graphs $A$ and $B$ from \Cref{def-corr-bern-net}, with $\Theta_{ij} = 0.1$, $1 \leq i \neq j \leq n$, $n = 300$ and $(s, \rho) = (0.9, 1)$.  Ideally, for each vertex $i$ in $A$, its true match $\pi^*(i)$ in $B$ should be closest to $i$, in terms of the $W_1$ distance.  In practice, this is not always true, even though the correlation parameter $\rho = 1$.  In \Cref{fig-EE-motive}, we plot the ranks of $\pi^*(i)$, for all $i$, among all their competitors.  The left panel includes all the vertices, and the right panel is a zoomed-in version of the top 50 of the left panel.  As we can see, the true ones do not always possess the smallest distance to their matches, but are among the smallest ones most of the cases.  

With the introduction of the parameter $d$, in terms of correctly matched pairs, \Cref{alg-edge-exp-dp} of course improves substantially over \Cref{alg-graph-matching}, which we will elaborate in \Cref{sec-numerical}.  The rationale behind is that in reality, adopting \Cref{alg-graph-matching} will nail down the matching to a small size of candidates.  If the requirements on accuracy are not to the individual level, then instead of matching each vertex to at most one vertex, one would pay the price of increasing the matching size in order to find the correct matching.  This is common in advertizing, for instance.  This also shares similarity with \cite{fishkind2012seeded}, where the output is a probability distribution attached to each vertex representing the probability of potential matches.  The output of \Cref{alg-edge-exp-dp} can be regarded as a uniform distribution over $d$ potential matches.

\subsection{Refinement}\label{sec-refine}

The key component of the degree profile graph matching algorithms in Bernoulli networks is discussed in Algorithms~\ref{alg-graph-matching} and \ref{alg-edge-exp-dp}.  In practice, \Cref{alg-graph-matching} suffers from the small matching size and unsatisfactory recovery rate, and the \Cref{alg-edge-exp-dp} can only provide a matching set for each vertex.  It is of question whether any additional steps can help to refine the matching result.  In this subsection, we discuss two refinement algorithms, focusing on preprocessing and post processing, respectively.

\subsubsection{Preprocessing}\label{sec-pre}
In practice, the high degree vertices have many neighbours and enjoy ample information for a successful matching. A natural idea is to first find such high degree vertices and their counterparts in the other graph, and then extend the matchings of high degree vertices only to matchings of all.  This can be done by finding vertices with degrees larger than a pre-specified threshold.

Based on this idea, we propose the seeded edge exploited graph matching algorithm in \Cref{alg-edge-exp-dp-pre}.  We first establish a collection of matches $\pi_0$ for the high degree vertices (degrees are at least $\tau_1$), the distances of which are the smallest among the pairs in consideration (distances are at most $\tau_2$).  The set of these vertices is called the \emph{seeds set}, $\mathcal{S}$.  Next, we calculate the similarity between $i \in A$ and $j \in B$ using $W_{ij} = \sum_{k \in \mathcal{S}} A_{ik} B_{j \pi_0(k)}$, the number of common neighbours between $i$ and $j$ based on $\pi_0$.  We then turn the similarity matrix $W$ to a bipartite adjacency matrix using the threshold $\tau_3$.   With the maximum bipartite matching, we find a one-to-one correspondence between $A$ and $B$ as $\pi_1$, so that the number of common neighbours is maximized.  Finally, we calculate the similarity again based on $\pi_1$, and find the matching set for each vertex as the vertices set with largest similarity.  Details can be found in \Cref{alg-edge-exp-dp-pre}.

\begin{algorithm}[htbp]
\begin{algorithmic}
	\INPUT $A \in \{0, 1\}^{n_A \times n_A}$, $B \in \{0, 1\}^{n_B \times n_B}$, $\tau_1, \tau_2, \tau_3 > 0$, a positive integer $d \geq 1$, 
	\For{$i = 1, \ldots, n_A$}
		\State $a_i \leftarrow \sum_{j = 1}^{n_A} A_{ij}$ 
	\EndFor
	\For{$i = 1, \ldots, n_B$}
		\State $b_i \leftarrow \sum_{j = 1}^{n_B} B_{ij}$ 
	\EndFor
	\State $W \leftarrow W(A, B)$ \Comment{\Cref{alg-0}}
	\State $\mathcal{S} = \{(i, \pi_0(i))\} \leftarrow \{(i, k): \, a_i \geq \tau_1, b_k \geq \tau_1, W_{ik} \leq \tau_2\}$
	\State $T_1, T_2 \leftarrow $ the collection of first and second coordinates of the members in $\mathcal{S}$
	\For{$i \in \{1, \ldots, n_A\} \setminus T_1$}
		\For{$k \in \{1, \ldots, n_B\} \setminus T_2$}
			\State $U_{ik} \leftarrow \mathbbm{1}\left\{\sum_{l \in T_1} A_{il}B_{k \pi_0(l)} \geq \tau_3\right\}$
		\EndFor
	\EndFor
	\State $\pi_1 \leftarrow \mathrm{MaxBipartiteMatching}(U)$
	\State $\pi \leftarrow \pi_0 \cup \pi_1$
	\For{$i = 1, \ldots, n_A$}
		\For{$j = 1, \ldots, n_B$}
			\State $W_{ij} \leftarrow \sum_{\{(k, \pi(k))\}} A_{ik} B_{j\pi(k)}$
		\EndFor
	\EndFor
	\State $Z \leftarrow 0_{n_A \times n_B}$, which is an all zero matrix
	\For{$i = 1, \ldots, n_A$}
		\State $\{(i, i_k)\}_{k = 1}^d \leftarrow $ the indices of the $d$ smallest entries in the $i$th row of $W$
		\State $(Z_{i i_k}, \, k = 1, \ldots, d) \leftarrow 1_d$
	\EndFor
	\OUTPUT $Z$
	\caption{Edge exploited degree profile graph matching with preprocessing \label{alg-edge-exp-dp-pre}}
\end{algorithmic}
\end{algorithm}

\begin{figure}
\begin{center}
\includegraphics[width = 0.6\textwidth]{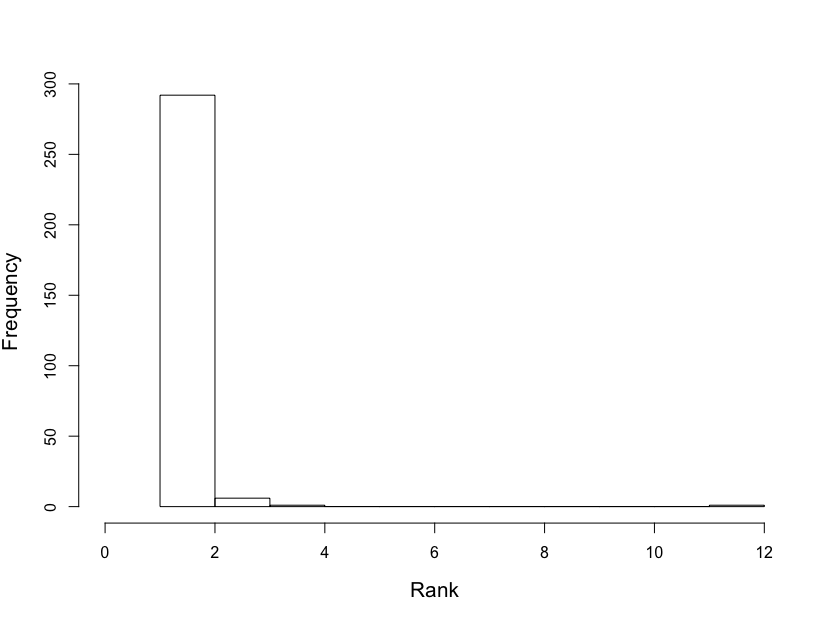}	
\end{center}
\caption{Histogram of the ranks of true pairs. Most true pairs have the smallest distance that will be chosen as the seeds. \label{fig-4}}
\end{figure}

In \Cref{alg-edge-exp-dp-pre}, there are three thresholds to find a proper original matching.  In practice, we conduct grid search to determine $\{\tau_1, \tau_2, \tau_3\}$.  For the two graphs, we calculate the degrees of all vertices and select 7 candidates for $\tau_1$, which correspond to the $i$-th quantile of vertices' degrees, $i \in \{0.5, 0.55, 0.6, 0.65, 0.7, 0.75, 0.8\}$.  Possible $\tau_2$ is chosen from the $j$-th, $j\in \{0.2, 0.25, 0.3, 0.35, 0.4, 0.45, 0.5\}$, quantile of the minimum distance between vertices in two graphs.  The best combination of $\tau_1$ and $\tau_2$ is supposed to give the largest collection of seeds. Having obtained seeds, the number of common neighbours are calculated between all pairs of vertices and denoted as $U_{ik}$.  The parameter $\tau_3$ is defined as the $\frac{n-1}{n}$-th quantile of $U_{ik}$, which guarantees that the number of nonzero elements in $U$ is approximately $n$.  If all possible combinations provided empty seed sets, \Cref{alg-edge-exp-dp} is summoned instead. 

The concept of seeded graph matching is used in other ways in the literature.  For instance, in \cite{lyzinski2014seeded} and \cite{lyzinski2015graph}, the seeds mean the information of some known vertices correspondence and seeded graph matching utilizes these known partial matching and includes them as constraints in the optimization.  In \cite{ding2018efficient}, the seeded degree profile graph matching starts with no known partial matchings and aims to refine Algorithm 1 in relatively dense graphs.  We would like to emphasize that this relatively dense regime studied there is even too sparse for spectral-based graph matching methods to perform well.

\Cref{alg-edge-exp-dp-pre} can be regarded as an edge exploited version of Algorithms~2 and 3 in \cite{ding2018efficient}.  As we have mentioned, the main task of this paper is to move beyond the perfectly-overlapping Erd\H{o}s--R\'enyi random graphs, therefore, in \Cref{alg-edge-exp-dp-pre}, we adopt an edge exploited version.  To motivate the preprocessing step, we alter the settings in \Cref{fig-EE-motive} slightly, by increasing the overlapping parameter $s$ from 0.9 to 0.99, which results in an easier problem.  In \Cref{fig-4}, we again exhibit the true ranks.  Different from \Cref{fig-EE-motive}, we can see that in this easier setting, almost all the true matchings are the ones with smallest distances.  A preprocessing step will return a set of true matching.

\subsubsection{Post processing}

The way to produce a seeds set in \Cref{alg-edge-exp-dp-pre} sheds light on the post-processing step.  With any preliminary graph matching result $\pi_t$ (this can be from either \Cref{alg-graph-matching} or \Cref{alg-edge-exp-dp}), we can define the similarity between $i \in A$ and $j \in B$ as 
	\[
		W_{ij} = \sum_{k \in A} \sum_{l \in \pi_0(k)} A_{i k} B_{j l},
	\]
	which is the number of common neighbours between $i$ and $j$ according to the matching $\pi_0$.   Based on the similarity matrix, we use maximum bipartite matching to maximize the number of common neighbours for the matched vertices. 
 
Now we rewrite the matching $\pi_t$ as $\Pi_t$, where $\Pi_t$ is an $n_A \times n_B$ permutation matrix with $(\Pi_t)_{ij} = 1$ if $j \in \pi_t(i)$ and 0 otherwise.  Given $\Pi_t$, The post processing step is to seek a refinement $\Pi_{t+1}$ satisfying 
	\begin{equation}\label{eq-iterate}
		\Pi_{t+1} \in \argmax_{\Pi_{n_A \times n_B} \mbox{\footnotesize is a permutation matrix}}\langle \Pi, A \Pi_t B \rangle.
	\end{equation}
	The intuition is to refine the result iteratively by optimizing this quadratic assignment problem.  Details are collected in \Cref{alg-edge-exp-dp-post}. 

\begin{algorithm}[htbp]
\begin{algorithmic}
	\INPUT $A \in \{0, 1\}^{n_A \times n_A}$, $B \in \{0, 1\}^{n_B \times n_B}$, $\tau > 0$, positive integers $d, n_{\mathrm{rep}} \geq 1$. 
	\State $\Pi_0 \leftarrow \mathrm{EE}(A, B, d)$	
	\State $\mathrm{FLAG}_0 \leftarrow 0_{n_A}$
	\For{$t = 1, \ldots, n_{\mathrm{rep}}$}
		\State $\Pi \leftarrow \argmax_{\Pi \in \mathcal{S}_n}\langle \Pi, A \Pi_0 B \rangle$
		\State $\mathrm{FLAG} \leftarrow \mathrm{FLAG}_{0} \cdot (\mathbbm{1}\{\Pi^{(i)} = \Pi_0^{(i)}\}, \, i = 1, \ldots, n_A)^{\top} + (\mathbbm{1}\{\Pi^{(i)} = \Pi_0^{(i)}\}, \, i = 1, \ldots, n_A)^{\top}$
		\State $\Pi_0 \leftarrow \Pi$; $\mathrm{FLAG}_0 \leftarrow \mathrm{FLAG} $
	\EndFor
	\State FLAG = $\mathbbm{1}\{\mathrm{FLAG} > \tau\}$
	\OUTPUT \{$\Pi_0$, FLAG\}.
	\caption{Edge exploited degree profile graph matching with post processing. \label{alg-edge-exp-dp-post}}
\end{algorithmic}
\end{algorithm}

In addition to the graph matching output $\Pi_0$, we also output a convergence indicator vector FLAG.  In practice, we have observed that the true matches usually reach the convergence and stay the same after a few iterations, while the false matches may keep changing in the iterations.  Instead of giving a guidance on the choice of $n_{\mathrm{rep}}$, we report the convergence indicators for each matching as a reference for the certainty about the matching. Default value for $\tau$ is $n_{\mathrm{rep}}/10$, which means for the final $10\%$ iterations, the matchings staying the same are regarded as ``converged''. 

The post processing algorithm in \Cref{alg-edge-exp-dp-post} is inspired by the iterative clean-up procedure proposed in \cite{ding2018efficient}.  \Cref{alg-edge-exp-dp-post} is shown to be numerically superior in more challenging setting and provides more information to improve the matching accuracy. 


\subsection{Graph matching in community-structured networks}\label{sec-comm}

Since most of the theoretically-justified graph matching algorithms are designed for perfectly-overlapping Erd\H{o}s--R\'enyi random graphs, including the degree profile graph matching, a natural question when we move beyond is whether to conduct graph matching directly on, say stochastic block models, or to conduct community detection first then match the graphs.  

Before we investigate this problem, we first state the community detection algorithm we adopt in this paper.  The spectral clustering on ratios-of-eigenvectors was proposed in \cite{jin2015fast} and detailed below for completeness.

\begin{algorithm}[htbp]
\begin{algorithmic}
	\INPUT a symmetric matrix $A \in \{0, 1\}^{n \times n}$, a positive integer $K \geq 2$
	\State $\{v_1, \ldots, v_K\} \leftarrow$ unit-length eigenvectors of $A$ corresponding to the $K$ leading singular values 
	\State $\{u_2, \ldots, u_K\} \leftarrow \{u_2/v_1, \ldots, u_K/v_1\}$
	\State $\{V_1, \ldots, V_K\} \leftarrow k$-means clustering based on the rows of $(v_1, u_2, \ldots, u_K)$ 
	\OUTPUT $\{V_1, \ldots, V_K\}$
	\caption{Spectral clustering on ratios-of-eigenvectors $\mathrm{SCORE}(A, K)$\label{alg-score} }
\end{algorithmic}
\end{algorithm}

\begin{algorithm}[htbp]
\begin{algorithmic}
	\INPUT $A \in \{0, 1\}^{n \times n}$, $B \in \{0, 1\}^{n \times n}$, positive integers $d, n_{\mathrm{rep}} \geq 1$, $K \geq 2$
	\State $(\{V_k^A\}_{k = 1}^K, \{V_k^B\}_{k = 1}^K) \leftarrow (\mathrm{SCORE}(A, K), \mathrm{SCORE}(B, K))$
	\For{$\mu \in \mathcal{S}_K$}
		\For{$k = 1, \ldots, K$}
			\State $\pi_k \leftarrow$ a graph matching result by matching $V_k^A$ and $V^B_{\mu(k)}$
		\EndFor
		\State $\pi_{\mu} \leftarrow \cup_{k = 1}^K \pi_k$		
	\EndFor
	\OUTPUT $\{\Pi_{\mu}, \mu \in \mathcal{S}_K\}$
	\caption{Degree profile graph matching with community detection \label{alg-score-match-0}}
\end{algorithmic}
\end{algorithm}

\begin{algorithm}[htbp]
\begin{algorithmic}
	\INPUT $A \in \{0, 1\}^{n \times n}$, $B \in \{0, 1\}^{n \times n}$, positive integers $d, n_{\mathrm{rep}} \geq 1$, $K \geq 2$
	\State $(\{V_k^A\}_{k = 1}^K, \{V_k^B\}_{k = 1}^K) \leftarrow (\mathrm{SCORE}(A, K), \mathrm{SCORE}(B, K))$
	\For{$\mu \in \mathcal{S}_K$}
		\For{$k = 1, \ldots, K$}
			\State $\pi_k \leftarrow$ a graph matching result by matching $V_k^A$ and $V^B_{\mu(k)}$
			\State $\mathrm{Eval}_k \leftarrow$ the corresponding evaluation of $\pi_k$
		\EndFor
		\State $\pi_{\mu} \leftarrow \cup_{k = 1}^K \pi_k$; $\mathrm{Eval}_{\mu} \leftarrow \sum_{k=1}^K \mathrm{Eval}_k$
	\EndFor
	\State $\pi = \pi_{\argmax_{\mu \in \mathcal{S}_K} \mathrm{Eval}_{\mu}}$
	\State $\Pi_0 \in \mathbb{R}^{n_A \times n_B} \leftarrow$ the matching matrix induced by $\pi$
	\State $\mathrm{FLAG}_0 \leftarrow 0_{n_A}$
	\For{$t = 1, \ldots, n_{\mathrm{rep}}$}
		\State $\Pi \leftarrow \argmax_{\Pi \in \mathcal{S}_n}\langle \Pi, A \Pi_0 B \rangle$
		\State $\mathrm{FLAG} \leftarrow \mathrm{FLAG}_{0} \cdot (\mathbbm{1}\{\Pi^{(i)} = \Pi_0^{(i)}\}, \, i = 1, \ldots, n_A)^{\top} + (\mathbbm{1}\{\Pi^{(i)} = \Pi_0^{(i)}\}, \, i = 1, \ldots, n_A)^{\top}$
		\State $\Pi_0 \leftarrow \Pi$; $\mathrm{FLAG}_0 \leftarrow \mathrm{FLAG} $
	\EndFor
	\State FLAG = $\mathbbm{1}\{\mathrm{FLAG} > \tau\}$
	\OUTPUT \{$\Pi$, FLAG\}
	\caption{Edge exploited degree profile graph matching with community detection \label{alg-score-match}}
\end{algorithmic}
\end{algorithm}

In \Cref{sec-SBM-sim}, we conduct a systematic investigation on the following two approaches:
	\begin{itemize}
	\item [(1)] first applying \Cref{alg-score}, then applying a graph matching algorithm within communities;
	\item [(2)]	directly applying a graph matching algorithm.
	\end{itemize}

There are various different community detection methods, even within the category of spectral-based methods.  As for the methods we have applied, there is no obvious differences between those based on \Cref{alg-score} and those based on other spectral clustering methods.  

As for the first approach, we further detail two algorithms listed in Algorithms~\ref{alg-score-match-0} and \ref{alg-score-match}.  In \Cref{alg-score-match-0}, we first apply \Cref{alg-score} and then use a certain graph matching method to match different communities.  Note that $\mathcal{S}_K$ is the collection of all possible permutations on $\{1, \ldots, K\}$.  We write \Cref{alg-score-match-0} in a generic and, in fact, incomplete way.  The output of \Cref{alg-score-match-0} has $K!$ many matching results.  \Cref{alg-score-match} can be regarded a post processing version of \Cref{alg-score-match-0} using the post processing method we introduced in \Cref{alg-edge-exp-dp-post}.    The quantity Eval in \Cref{alg-score-match} is short for evaluation, which is algorithm-specific.  For instance, if the graph matching algorithm used thereof is chosen to be \Cref{alg-graph-matching} or \Cref{alg-edge-exp-dp-post}, then Eval can be taken as the number of matched vertices or the number of converged vertices, respectively.

We now come back to investigate the choice between approaches (1) and (2).  The evaluation is twofold: the theoretical limits and the violation to the theoretical guarantees of the graph matching methods.

We first resort to the theoretical limits of Algorithms~\ref{alg-graph-matching} and \ref{alg-score}.  It is established \citep[see e.g.][Theorem~2.2]{rohe2011spectral} that to ensure the misclustered nodes are consisted of a vanishing ratio of all the nodes, the entries in $\Theta$ defined in \Cref{def-bern-net} are at least of order $\log^{-1/2}(n)$, which is a much stronger condition than the ones required in \Cref{alg-graph-matching}.  For instance, in order to achieve a perfect matching with high probability in two perfect overlapped correlated Erd\H{o}s--R\'enyi random graphs, the required lower bound on the Erd\H{o}s--R\'enyi parameter is of order $\log^2(n)/n$.  This is to say, in terms of the order of $\|\Theta\|_{\infty}$, the limit of approach (1) is at least $\log^{-1/2}(n)$, and $\log^2(n)/n$ in (2).  However, we should bear in mind that the $\log^2(n)/n$ is established for Erd\H{o}s--R\'enyi random graphs but not for stochastic block models.

In terms of the violations of the theoretical guarantees provided in \cite{ding2018efficient}, we first state the rationale behind the approach (1).  Since \Cref{alg-graph-matching} is only theoretically justified on correlated Erd\H{o}s--R\'enyi random graphs, it might be helpful to conduct community detection first to reduce a stochastic block model graph matching problem to a patially-overlapping Erd\H{o}s--R\'enyi one.  In fact, we cannot guarantee that with probability tending to 1, there is no misclustered vertex.  This means even if we are in a regime where the community detection is strongly consistent, the resulting community may still contain misclustered vertices.  The matching conducted in the approach (1) is a graph matching over partially-overlapping graphs.

\section{Simulation analysis}\label{sec-numerical}

In this section, we conduct a thorough simulation analysis on the numerical performances of the algorithms proposed in \Cref{sec-methods}.  For notational simplicity, we refer to Algorithms~\ref{alg-graph-matching}, \ref{alg-edge-exp-dp}, \ref{alg-edge-exp-dp-pre} and \ref{alg-edge-exp-dp-post} as DP (degree profile), EE (edge exploited version), EE-pre (preprocessing, EE-) and EE-post (post processing, EE+), respectively.  We will see that our proposed methods can perform well in challenging situations for partially-overlapping graphs and for stochastic block models.

\subsection{Partially-overlapping correlated Erd\H{o}s--R\'enyi random graphs}\label{sec-sim-ER}

In this subsection, we consider graph matching in patially-overlapping correlated Erd\H{o}s--R\'enyi random graphs. 

The simulation settings involve the following parameters: (i) the network size $n = 300$, (ii) the connection probability $q \in \{0.10, 0.05\}$, and (iii) $(\rho, s) \in \{(1, 1), (0.95, 0.98), (0.9, 0.95)\}$, where $\rho$ and $s$ are the correlation and overlapping parameters, respectively.  Each setting is repeated 50 times.

As for the tuning parameters used in the algorithms, we let $d \in \{10, 30\}$, where $d$ is the tuning parameter for the edge exploited step in Algorithms~\ref{alg-edge-exp-dp}, \ref{alg-edge-exp-dp-pre} and \ref{alg-edge-exp-dp-post}.  The tuning parameters required in \Cref{alg-edge-exp-dp-pre} are generated automatically based on the grid search method we introduced in \Cref{sec-pre}.

The methods we adopt are DP, EE, EE-pre and EE-post.  The performances are evaluated by the recovery rate over all nodes that have counterparts in the other graph. Note that we allow for partially-overlapping graphs, hence not every single node has a counterpart in the other graph, and the number of these overlapping nodes is usually smaller than the number of nodes in a single graph.

\begin{figure}
\centering
\includegraphics[width = \textwidth]{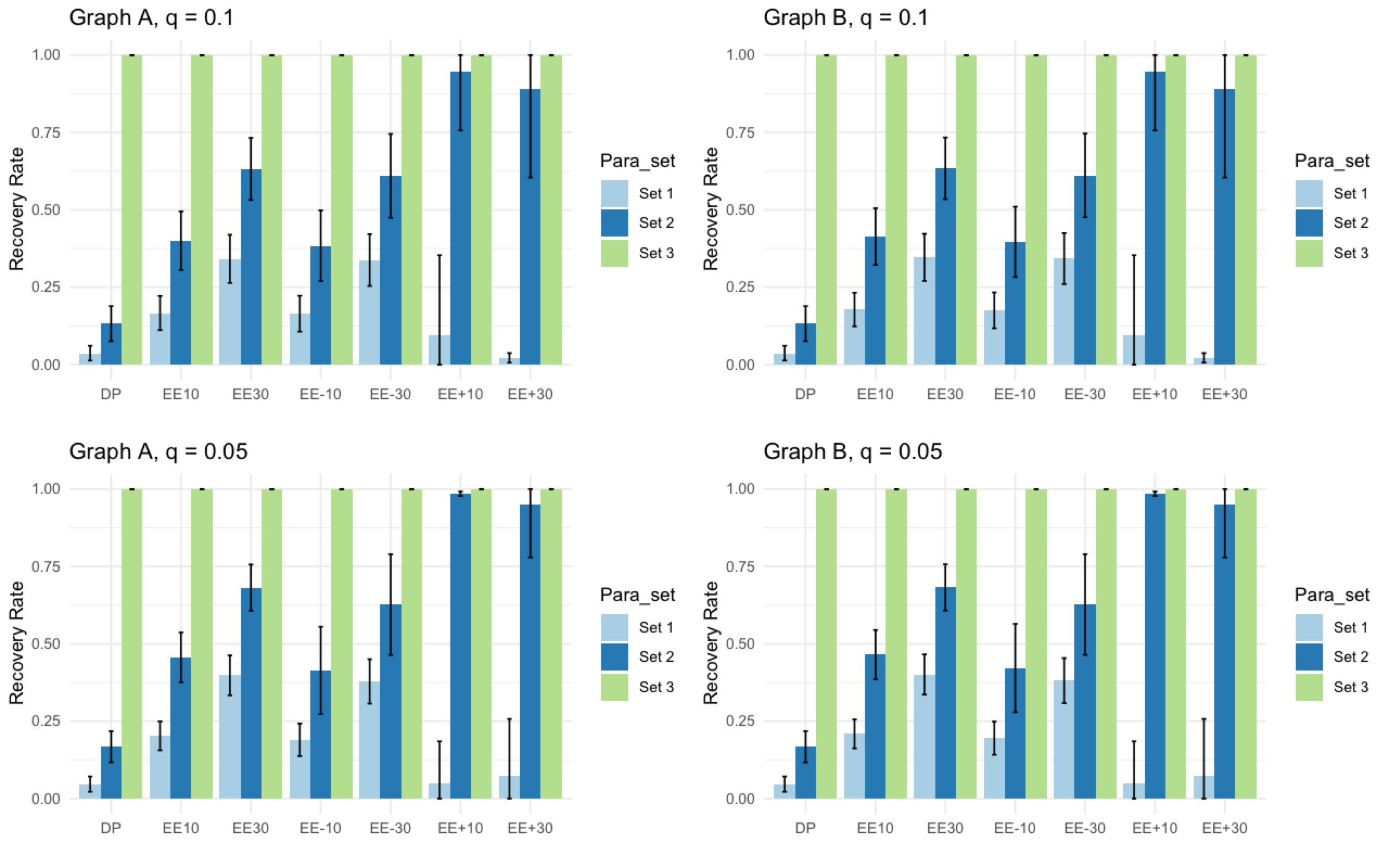}
\caption{Simulation results in \Cref{sec-sim-ER}.  All are the means and standard errors.  The parameter settings are: Set 1, $(\rho, s) = (0.9, 0.95)$; Set 2, $(\rho, s) = (0.95, 0.98)$;  Set 3, $(\rho, s) = (1, 1)$.  The methods are: DP, \Cref{alg-graph-matching}; EE10, \Cref{alg-edge-exp-dp} with $d = 10$; EE30, \Cref{alg-edge-exp-dp} with $d = 30$; EE-10, \Cref{alg-edge-exp-dp-pre} with $d = 10$; EE-30, \Cref{alg-edge-exp-dp-pre} with $d = 30$; EE+10, \Cref{alg-edge-exp-dp-post} with $d = 10$; EE+10, \Cref{alg-edge-exp-dp-post} with $d = 30$. \label{fig-ER-sim}}	
\end{figure}

The results are collected in \Cref{fig-ER-sim}.   Since our methods are asymmetry for the two graphs, we present the recovery rates for each graph separately.   Despite the asymmetry, the difference between graphs are negligible.  

The three parameter settings are in difficulty decreasing order.  In Setting 1, $(\rho, s) = (0.9, 0.95)$, in terms of the recovery rate, EE algorithms are the best.  This is not a surprise, since they are the only ones allowing for matching one node to multiple nodes.  Even so, the recovery rate is still below half.  In Setting 3, $(\rho, s) = (1, 1)$, the two graphs are identical.  All algorithms behave well.  In Setting 2, $(\rho, s) = (0.95, 0.98)$, we can see that EE-post dominantly outperformed all the other methods, even though EE algorithms allow for multiple matching while EE-post only for single matching. 

Among all settings, EE-post with $d = 30$ is similar or worse than the case $d = 10$. It suggests a small tuning parameter $d$ for successful results. 
Besides the recovery rate, there is also a convergence parameter FLAG for EE-post algorithm. Interestingly, if  we roughly regard the iterations with $\sum \mbox{FLAG}_i > n/2$ as iterations that EE-post succeeds, then EE-post has recovery rate around 0.9 for all the successful iterations, and approximately 0 for others. The convergence indicator provides supporting information to decide whether the matching is reliable or not. 

\subsection{Correlated stochastic block models}\label{sec-SBM-sim}

In this subsection, we consider graph matching in correlated stochastic block models.  Different from \Cref{sec-sim-ER}, we only consider perfectly-overlapping graphs. In \Cref{sec-comm}, we have discussed that different theoretical limits for community detection and graph matching may induce misclustered nodes and hence partially-overlapping graphs to match. 

The simulation settings involve the following parameters: (i) the network size $n = 1000$, (ii) the number of communities $K = 2$, (iii) the within communities probability $q \in \{0.10, 0.05\}$ and the between communities probability $q/2$, and (iv) the probability of keeping an edge from the parent graph $\rho \in \{0.95, 0.93, 0.9\}$.  Each setting is repeated 10 times.

As for the tuning parameters used in the algorithms, we let $d \in \{10, 50\}$, where $d$ is the tuning parameter for the edge exploited step in Algorithms~\ref{alg-edge-exp-dp}, \ref{alg-edge-exp-dp-pre} and \ref{alg-edge-exp-dp-post}.  The tuning parameters required in \Cref{alg-edge-exp-dp-pre} are generated automatically based on the grid search method we introduced in \Cref{sec-pre}.

In this scenario, we compare results from six different methods. (i) \Cref{alg-graph-matching}, (ii) \Cref{alg-edge-exp-dp-post}, (iii) \Cref{alg-score-match-0} with \Cref{alg-graph-matching}, (iv) \Cref{alg-score-match-0} with  \Cref{alg-edge-exp-dp-post}, (v) \Cref{alg-score-match} with \Cref{alg-graph-matching} and (vi) \Cref{alg-score-match} with \Cref{alg-edge-exp-dp-post}.  In \Cref{sec-comm} we have mentioned that the output of Algorithms~\ref{alg-score-match-0} and \ref{alg-score-match} are not necessarily unique.  In (iii) and (v), we choose the permutations of the communities which return more matchings.  In (iv) and (vi), we report the ones with larger converging matchings.  The measurements we adopt here are similar to those in \Cref{sec-sim-ER}, except that in this section, we do not report the results for graphs $A$ and $B$ separately.  Since we let $s = 1$ and the algorithms we evaluate only report at most one matching, the recovery results for graphs $A$ and $B$ are identical. 

\begin{figure}[htbp]
\centering
\includegraphics[width = \textwidth]{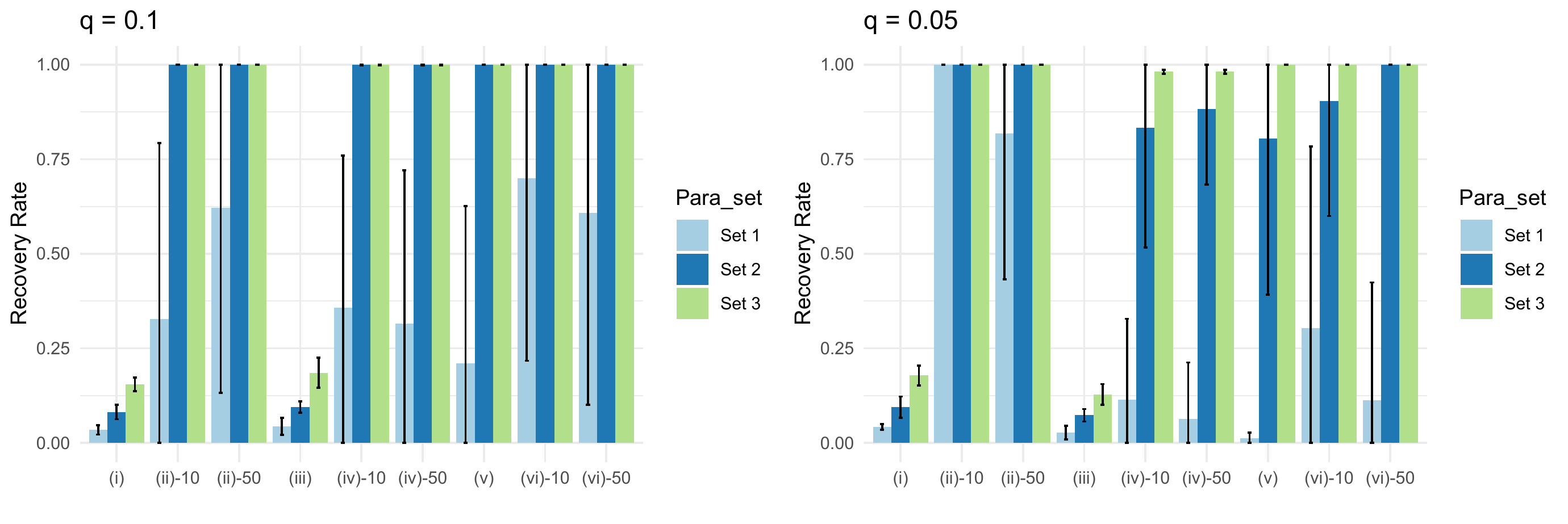}
\caption{Simulation results in \Cref{sec-SBM-sim}.  All are the means and standard errors.  The parameter settings are: Set 1, $\rho = 0.9$; Set 2, $\rho = 0.93$; Set 3, $\rho = 0.95$.  The methods are: (i), \Cref{alg-graph-matching}; (ii)-10, \Cref{alg-edge-exp-dp-post} with $d = 10$; (ii)-50, \Cref{alg-edge-exp-dp-post} with $d = 50$; (iii), \Cref{alg-score-match-0} with \Cref{alg-graph-matching}; (iv)-10 \Cref{alg-score-match-0} with  \Cref{alg-edge-exp-dp-post}, $d = 10$; (iv)-50 \Cref{alg-score-match-0} with  \Cref{alg-edge-exp-dp-post}, $d = 50$; (v), \Cref{alg-score-match} with \Cref{alg-graph-matching}; (vi)-10, \Cref{alg-score-match} with \Cref{alg-edge-exp-dp-post}, $d = 10$; (vi)-50, \Cref{alg-score-match} with \Cref{alg-edge-exp-dp-post}, $d = 50$. \label{fig-SBM-sim}}
\end{figure}

Setting 1 is the most difficult one.  For this setting, EE-post methods can actually achieve almost perfect recovery for relatively sparse graphs ($q = 0.05$, right column panels). Another interesting thing to notice in Setting 1 is that, EE-post with $d = 10$ can perform better in the sparse graphs while EE-post with $d = 50$ performs better in the dense graphs. It may indicate a choice of small $d$ for sparse graphs in practice. 
In Setting 3, all algorithms perform well except (i) and (iii), both of which are based on \Cref{alg-graph-matching}. 

In order to answer the question that if one should do community detection before matching two stochastic block models, we recall that algorithm (ii) is to directly match stochastic block models, (iv) is to conduct EE-post on estimated communities and (vi) is to conduct EE-post on the estimated communities first and then the whole graph.  The comparable settings for this matter are Settings 1 and 2.  We can see that in the denser graphs (left column panel), conducting EE-post on both the estimated communities and the whole graph perform best.  In the sparser graphs (right column panel), directly matching graphs perform best.  This is to some extent expected, since the success of community detection relies on more stringent density requirements than the degree profile algorithms.

\section{Real data}\label{sec-data}

In this section, we conduct analysis on two real datasets and focus on the performance of \Cref{alg-graph-matching}, \Cref{alg-edge-exp-dp} and \Cref{alg-edge-exp-dp-post}.  

\subsection{Coauthor dataset}\label{sec-coauthor}
In this section, we analyse the coauthorship dataset, which is originally studied in \cite{ji2016coauthorship}, to find the co-authorship patterns between statisticians according to the publications in the Annals of Statistics (AoS), Biometrika, Journal of American Statistician Association (JASA) and Journal of Royal Statistical Society, Series B (JRSSB), during the period Years 2003-2012. 

For any three distinct journals $A$, $B$ and $C$ chosen from the above mentioned four journals, we construct two networks. One network is formed by the authors who published papers in $A$ and/or $B$, namely $A \cup B$.  The other network is formed by the authors who published papers in $A$ and/or $C$, namely $A \cup C$.   In $A \cup B$($A \cup C$), a node is an author who have published in $A \cup B$($A \cup C$), and an edge indicates that the corresponding two authors have at least one coauthored paper published in $A \cup B$($A \cup C$).  This construction provides partially-overlapping networks.  

Since the networks contain isolated nodes and pairs, which provide little information for graph matching, we preprocess the two networks as follows.  An author is kept only when they has common coauthors in both $A\cup B$ and $A\cup C$.  We then extract the giant components of these two networks respectively.  The resulting giant components are the final networks we work on.  Note that, the sizes of the giant components are about half of the original networks, and the final two networks have different size. 

As for EE and EE-post, we let $d = 5$ and $n_{\mathrm{rep}} = 50$.  In EE-post, we let $\tau = 5$.  It means we consider a matching as ``converged matching'' when the matching stays the same for at least last 5 iterations.  We introduce this new notion in the real data analysis, since our methods perform well without this additional criteria in the simulated data.

\begin{table}[htbp]
\caption{Dataset sizes of those studied in \Cref{sec-coauthor}.  Size $A\cup B$: the size of the giant component in processed $A \cup B$; Size $A \cup C$: the size of the giant component in processed $A \cup C$; Size Overlap: the number of overlapping nodes of the two networks.}
\label{table:coauthor}
\begin{center}
\begin{tabular}{rrrrr}
	Data A&  B vs C & \multicolumn{3}{c}{Size} \\
	&  & $A \cup B$ & $A \cup C$ & Overlap \\
	\hline
	\multirow{3}{*}{AoS} & Biometrika vs JASA & 682 & 658 & 593 \\
	 & JRSSB vs JASA &507 & 687 & 469 \\
	  & Biometrika vs JRSSB &610 & 458 & 451 \\
	  \multirow{3}{*}{JASA} & AoS vs JRSSB & 1010 & 984 & 870 \\
	  & AoS vs Biometrika & 1004 & 1023 & 890 \\
	  & JRSSB vs Biometrika & 984 & 1016 & 877 \\
	  \multirow{3}{*}{JRSSB} & JASA vs Biometrika &    419   &  386   &  323 \\
	  & AoS vs Biometrika &  258   &  351   &  232  \\
	  & AoS vs JASA &  272  &   369   &  222  \\
	  \multirow{3}{*}{Biometrika} & JASA \& JRSSB &589&  543 & 499 \\
	  & JASA vs AoS & 569  & 512 & 416 \\
	  & JRSSB vs AoS & 518  &   474  &   401 \\	  
\end{tabular}
\end{center}
\end{table}

\begin{figure}
\centering	
\includegraphics[width = \textwidth]{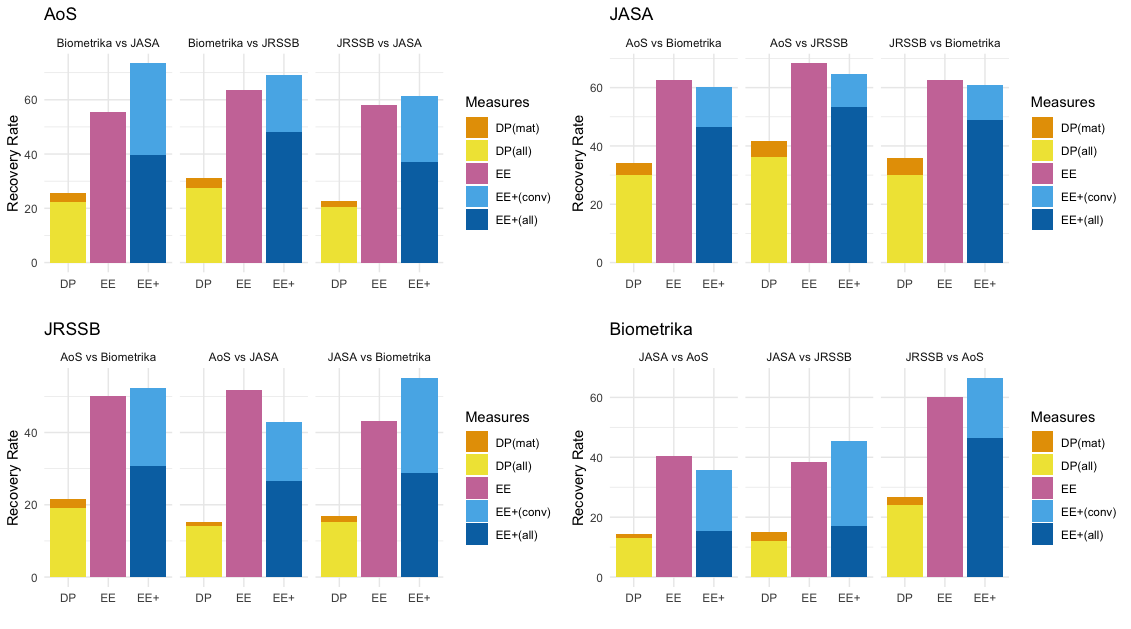}
\caption{Results in \Cref{sec-coauthor}.  Each panel corresponds a different Journal $A$, as indicated at the top-left corner of the panel.  Every three consecutive bars in a panel correspond to a different choice of Journals $B$ and $C$, as indicated at the top of the bars.  The metrics are: DP(mat), the recovery rate of \Cref{alg-graph-matching} over all matched pairs; DP(all), the recovery rate of \Cref{alg-graph-matching} over all nodes; EE, the recovery rate of \Cref{alg-edge-exp-dp} in terms of all nodes; EE+(all), the recovery rate of \Cref{alg-edge-exp-dp-post} over all nodes; EE+(conv), the recovery rate of \Cref{alg-edge-exp-dp-post} over all converged pairs. \label{fig-coauthor}}
\end{figure}

The detailed network sizes exhibited in \Cref{table:coauthor}.  Note that, no matter which combination of journals is considered, the corresponding pairs of networks are partially overlapping.  In fact, the sizes of overlapped nodes are much smaller than that of networks. 

In \Cref{fig-coauthor}, we depict the recovery rates over five different metrics.  DP(all) is the recovery rate of \Cref{alg-graph-matching} over all nodes, and it is smaller than DP(mat), which is the recovery rate of \Cref{alg-graph-matching} over all matched nodes.  Apparently, DP(mat) is alway larger than DP(all), so in \Cref{fig-coauthor}, we stack the differences between these two on top of DP(all).  The larger the differences are, the smaller the matched nodes ratios are.  As for \Cref{alg-edge-exp-dp-post}, we also consider two metrics, EE+(all) -- the recovery rate in terms of all nodes, and EE+(conv) -- the recovery rate in terms of converged nodes.  In \Cref{fig-coauthor}, we also stack the difference between EE+(conv) and EE+(all) on top of EE+(all).

A fair comparison is to compare EE+(conv) with DP(mat), and to compare EE+(all) with DP(all).  As we can see, EE-post consistently and substantially outperform DP in all aspects.  In particular, EE+(conv) shows an even more prominent improvement, which suggests that, for real data where the underlying truth is unknown and the matching accuracy is of concern, we can use the converged matchings of EE-post as a reliable matching. 

To provide more insights of EE-post methods, we examine three specific authors in the coauthor dataset.  We use the dataset AoS $\cup$ Biometrika and AoS $\cup$ JASA for illustration.
	\begin{itemize}
	\item Converged and correctly matched.	An example of this category is Author 60.  It has in total three coauthors in the dataset concerned, and all these three coauthors occur in the AoS.  This suggests that Author 60 has the same size of neighbourhood in AoS $\cup$ Biometrika and AoS $\cup$ JASA.  In addition, at least one of these three neighbours is correctly matched.  These two facts provide ample information for graph matching, and result in Author 60 being a converged node in EE-post algorithm and is correctly matched.
	\item Converged but wrongly matched.  An example of this category is Author 222.  It has zero coauthor in AoS, four in JASA and four in Biometrika.  The intersection of it's JASA and Biometrika collaborators sets is of size three.  In terms of graph matching Author 222, the interference signal comes from Author 655, who share two coauthors with Author 222 and who is wrongly matched to Author 222.  A relatively large number of coauthors leads to the convergence, while the interference signal results in a wrong match.
	\item Correctly matched but not converged.  An example of this category is Author 115.  Note that Author 115 has five coauthors in the dataset concerned, but only one of these five neighbours is correctly matched.  This causes that in the iterations, the matching of Author 115 is not stable, but one possible matching is correct due to the relatively large number of neighbours.  This example also sheds light on the rationale of adopting EE with $d > 1$, when one can afford a multiple matching storage.
	\end{itemize}

\subsection{Zebrafish dataset}\label{sec-zebrafish}
In this section, we analyse a zebrafish neuronal activity dataset.  This dataset is originally acquired and processed in \cite{prevedel2014simultaneous} and is a time series of whole-brain zebrafish neuronal activity.  We follow the preprocessing routine conducted in \cite{lyzinski2017fast} and subtract a slice of neuronal activity network which is in fact the sample correlation matrix in a small window of time.  This can be regarded as the adjacency matrix of a weighted undirected network, with 5105 nodes.  The further analysis conducted in this section is based on thresholding the entries of this correlation matrix $R$ to provide adjacency matrices in $\{0, 1\}^{5105\times 5105}$.

We conduct two sets of simulation based on this dataset.  One is to match graphs generated from two different thresholds and the other is based on the same thresholds.  To be specific, in the different thresholds setting, we first use threshold $t_1 \in \{0.5, 0.6, 0.7\}$ to produce a matrix $A_1 \in \{0, 1\}^{5105 \times 5105}$, by letting $(A_1)_{ij} = \mathbbm{1}\{R_{ij} \geq t_1\}$, and use $t_2 = t_1 + 0.1$ to produce $B_1 \in \{0, 1\}^{5105 \times 5105}$.  For each of $A_1$ and $B_1$, we then subtract the leading principal sub-matrix $A_2, B_2 \in \{0, 1\}^{m \times m}$, $m \in \{100, 300, 1000\}$.  Finally, for each node in $A_2$($B_2$), we independently keep it with probability $s \in \{0.95, 0.97\}$ to produce $A_3$($B_3$), and output $A$($B$) by deleting isolated nodes.  When matching $A$ and $B$, we also randomly permute the nodes in $B$ to increase difficulty.  In the same threshold setting, we let $A_1 = B_1$ using the same threshold $t \in \{0.5, 0.6, 0.7\}$, and follow the rest of the procedures as those in the different threshold scenarios.  It is worth mentioning that, in the different threshold scenario, the higher threshold graph is a sub-graph of the lower threshold graph; in the same threshold scenario, we have that $\rho = 1$.

Each combination of the parameters mentioned above is repeated 10 times.  In particular, in the same threshold setting, the repetitions are conducted by permuting the nodes 10 times.

\begin{figure}
\centering	
\includegraphics[width = \textwidth]{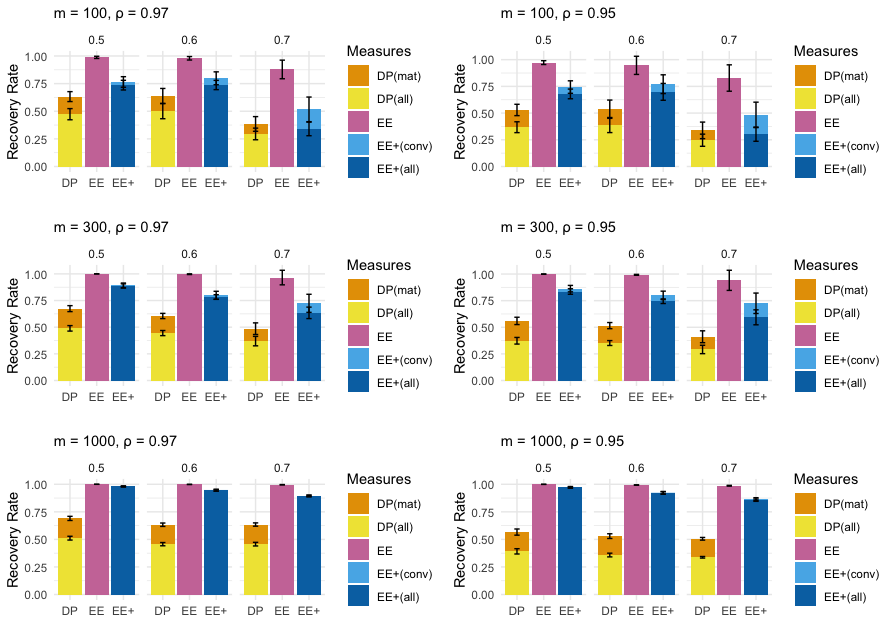}
\caption{Results in \Cref{sec-zebrafish}, different thresholds scenarios.  Each bar indicates the mean and standard error of a certain metric.  The size $m$ and the correlation parameter $\rho$ are indicated in each panel.  In each panel, every three consecutive bars represent metric values for a threshold value, indicated at the top.  The metrics are: DP(mat), the recovery rate of \Cref{alg-graph-matching} over all matched pairs; DP(all), the recovery rate of \Cref{alg-graph-matching} over all nodes; EE, the recovery rate of \Cref{alg-edge-exp-dp} in terms of all nodes; EE+(all), the recovery rate of \Cref{alg-edge-exp-dp-post} over all nodes; EE+(conv), the recovery rate of \Cref{alg-edge-exp-dp-post} over all converged pairs. \label{fig-zebra-diff}}
\end{figure}

\begin{figure}
\centering	
\includegraphics[width = \textwidth]{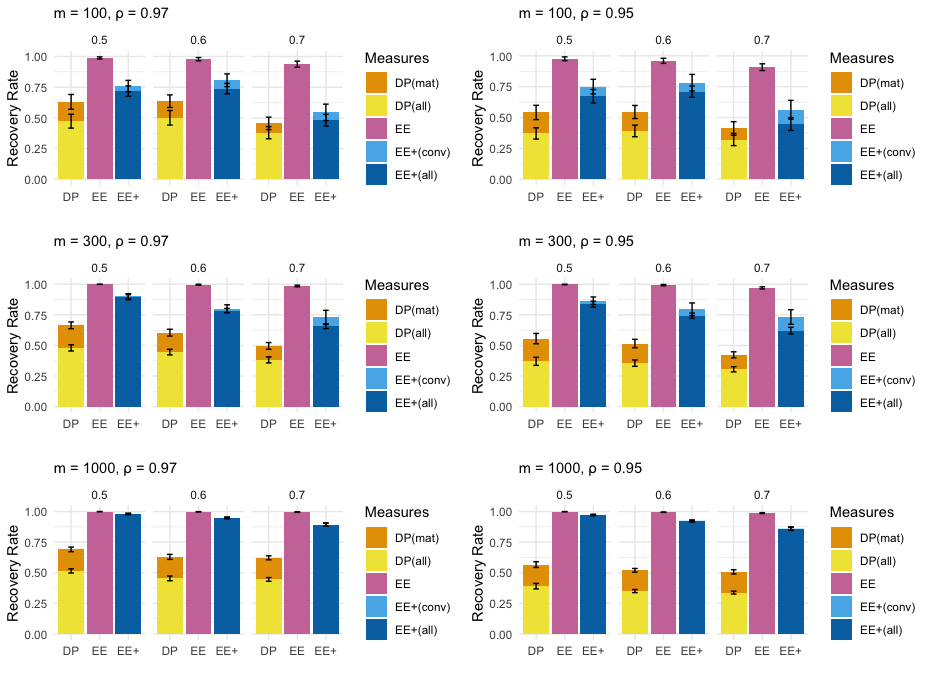}
\caption{Results in \Cref{sec-zebrafish}, same thresholds scenarios.  Each bar indicates the mean and standard error of a certain metric.  The size $m$ and the correlation parameter $\rho$ are indicated in each panel.  In each panel, every three consecutive bars represent metric values for a threshold value, indicated at the top.  The metrics are: DP(mat), the recovery rate of \Cref{alg-graph-matching} over all matched pairs; DP(all), the recovery rate of \Cref{alg-graph-matching} over all nodes; EE, the recovery rate of \Cref{alg-edge-exp-dp} in terms of all nodes; EE+(all), the recovery rate of \Cref{alg-edge-exp-dp-post} over all nodes; EE+(conv), the recovery rate of \Cref{alg-edge-exp-dp-post} over all converged pairs. \label{fig-zebra-same}}
\end{figure}

The numerical results are depicted in Figures~\ref{fig-zebra-diff} and \ref{fig-zebra-same}, for the different and same thresholds settings, respectively.  As for \Cref{alg-graph-matching}, we calculate the recovery rates over all nodes, DP(all) and matched nodes, DP(mat), separately.  Since DP(mat) is always larger than DP(all), we stack the difference between these two rates on top of the DP(all) in the figures.  As for \Cref{alg-edge-exp-dp-post}, we calculate the recovery rates over all nodes, EE+(all) and converged nodes, EE+(conv), separately.  For the same reasons as stated for \Cref{alg-graph-matching}, we stack the two bars in one in each panel in the figures.  

Generally speaking, as the thresholds increase, all the performances deteriorate, since the networks become sparser and the matching problems become harder.   The two scenarios, different and same thresholds, show very similar information, and in most of cases, all methods perform slightly better in the same threshold scenario.  It is interesting to see that EE has almost full recovery in most settings, even though this is based on real datasets.  Since the convergence rates of EE-post are high across all settings, the recovery rates of EE-post in two different metrics are comparable.  Overall, EE and EE-post outperform DP.  We would like to point out, as the network size increases, EE and EE-post improve their performances, while DP deteriorates.  This further suggests that in reality, EE-type methods are preferable over the original DP algorithm.

\section{Discussions}\label{sec-diss}

In this paper, we investigated the extensions of the degree profile graph matching in perfectly-overlapping Erd\H{o}s--R\'enyi random graphs.  The extensions include partially-overlapping graphs matching and stochastic block model graph matching.  We proposed the edge exploited graph matching algorithm and its variants, and conducted thorough numerical experiments to evaluate their performances.

In \Cref{def-bern-net}, we focused on simple graphs, i.e.~there are no multiple edges between any give pair of nodes.  The extension to multiple edge networks is straightforward, since all our current methods are based on counting the edges.  Other possible extensions include graph matching on directed graphs and the theoretical guarantees associated.  We will leave these for future work.





\bibliographystyle{Chicago}
\bibliography{ref}
\end{document}